\documentclass[prx,twocolumn, superscriptaddress, svgnames]{revtex4-2}
\usepackage[T1]{fontenc} 
\usepackage{tikz,tikzscale, relsize, pgfplots, xcolor, amssymb, amsmath}
\usetikzlibrary{backgrounds, arrows, calc, decorations, positioning, decorations.pathmorphing,decorations.pathreplacing, 	decorations.markings, fixedpointarithmetic, fit, patterns, decorations.text, backgrounds, matrix, plotmarks, spy, 3d, shapes.multipart, angles, quotes}
\usepgfplotslibrary{fillbetween}
\pgfplotsset{compat=1.17}
\usetikzlibrary{external}
\usepackage{graphicx}
\usepackage{ragged2e}

\usepackage{array}

\usepackage{arrayjobx}
\graphicspath{{./}}
\DeclareGraphicsExtensions{.pdf,.png,.jpg}

\usepackage{soul}

\usetikzlibrary{arrows.meta}

\begin{document}
\title{Space Group Symmetry of Chiral Fe-deficient van der Waals Magnet $\text{Fe}_{\text{3-x}}\text{GeTe}_{\text{2}}$ Probed by Convergent Beam Electron Diffraction}

\author{O. Zaiets}
\email{o.zaiets@ifw-dresden.de}
\affiliation{Leibniz Institute for Solid State and Materials Research Dresden, Helmholtzstraße 20, 01069 Dresden, Germany}
\affiliation{Institute of Solid State and Materials Physics, TU Dresden, Haeckelstraße 3, 01069 Dresden, Germany}
\author{S. Subakti}
\affiliation{Leibniz Institute for Solid State and Materials Research Dresden, Helmholtzstraße 20, 01069 Dresden, Germany}
\author{D. Wolf}
\affiliation{Leibniz Institute for Solid State and Materials Research Dresden, Helmholtzstraße 20, 01069 Dresden, Germany}
\author{J. Steinweh}
\affiliation{Leibniz Institute for Solid State and Materials Research Dresden, Helmholtzstraße 20, 01069 Dresden, Germany}
\affiliation{Institute of Solid State and Materials Physics, TU Dresden, Haeckelstraße 3, 01069 Dresden, Germany}
\author{S. Parkin}
\affiliation{Department for Nano-Systems from Ions, Spins, and Electrons (NISE), Max Planck Institute of Microstructure Physics, Weinberg 2, 06120 Halle(Saale), Germany}
\author{A. Lubk}
\email{a.lubk@ifw-dresden.de}
\affiliation{Leibniz Institute for Solid State and Materials Research Dresden, Helmholtzstraße 20, 01069 Dresden, Germany}
\affiliation{Institute of Solid State and Materials Physics, TU Dresden, Haeckelstraße 3, 01069 Dresden, Germany}
\affiliation{W\"urzburg--Dresden Cluster of Excellence ct.qmat, TU Dresden, 01062 Dresden, Germany}

\begin{abstract}
The crystal symmetry of Fe-deficient $\text{Fe}_{\text{3-x}}\text{GeTe}_{\text{2}}$ at room temperature has been investigated by a combination of selected-area electron diffraction (SAED) and convergent-beam electron diffraction (CBED). By symmetry analysis of CBED patterns along different zone axis, the space group of $\text{Fe}_{\text{3-x}}\text{GeTe}_{\text{2}}$ at room-temperature has been identified as $P6_3mc$ (No.186), which derives from the high-symmetry parent system $\text{Fe}_{\text{3}}\text{GeTe}_{\text{2}}$ ($P6_3/mmc$) by breaking the mirror symmetry along the 6-fold rotation axis. The $P3m1$ (No.156) space group previously reported for $\text{Fe}_{\text{3-x}}\text{GeTe}_{\text{2}}$ is a subgroup of $P6_3mc$ suggesting further possible symmetry breaks in this non-stoichiometric system. 
\end{abstract}

\maketitle
\section{Introduction}

The two-dimensional (2D) van der Waals (vdWs) magnets with general formula $\text{Fe}_{\text{n}}\text{GeTe}_{\text{2}}$ (n = 3, 4, 5) have recently emerged not only as a promising platform for unconventional physics, but also for technological applications in spintronics~\cite{Zhang23}. Among this family of compounds, the Fe-deficient $\text{Fe}_{\text{3-x}}\text{GeTe}_{\text{2}}$ system has attracted particular attention due to the presence of a metallic ferromagnetic (FM) state on a layered hexagonal lattice (centrosymmetric space group $P6_3/mmc$ (No.194) at room temperature with unit cell parameters $a\approx3.99$~\r{A}, $c\approx16.33$~\r{A}), where $\text{Fe}_{\text{3-x}}\text{Ge}$ slabs are sandwiched between two vdWs bonded Te layers~\cite{Zhuang16,May16,Milosavljevic19,Yang22,Deiseroth06}. Here, the quasi-2D nature in combination with sizeable spin-orbit coupling leads to strong magneto-crystalline anisotropy and a rather high Curie temperature (Tc $\sim220$~K)~\cite{Liu20}. Consequently, deliberately engineering cation vacancies and cation element substitution in $\text{Fe}_{\text{3-x}}\text{GeTe}_{\text{2}}$ opens a way to tailor the magnetic properties~\cite{May16}.

The existence of N\'eel-type skyrmions in bulk $\text{Fe}_{\text{3-x}}\text{GeTe}_{\text{2}}$ \cite{Chakraborty22, Liu2023} and their possible manipulation via electric current pulses predestines this material system as prominent building blocks for spintronic devices~\cite{LeonBrito16, Lopes21, Eom23}. Here, literature offers different mechanism for the existence of N\'eel-type skyrmions in this nominally centrosymmetric system. Xu et al. argue that specific forms of fourth-order interaction can stabilize skyrmions even in centrosymmetric $\text{Fe}_{\text{3-x}}\text{GeTe}_{\text{2}}$~\cite{Xu22}. A second possible mechanism is the breaking of inversion symmetry of the high-symmetry phase (point group $6/mmm$) to non-centrosymmetric point groups of type $C_{6v}$ ($6mm$), $C_{3v}$ ($3m$), or $C_{2v}$ ($mm2$) allowing for a non-zero antisymmetric exchange (Dzyaloshinkii-Moriya) interaction (DMI)~\cite{Birch22} that favors N\'eel-type spin modulations. Indeed, Chakraborty et al. proposed the non-centrosymmetric trigonal $P3m1$ (No.156)~\cite{Chakraborty22} space group (point group symmetry $3m$) as proper symmetry of $\text{Fe}_{\text{3-x}}\text{GeTe}_{\text{2}}$, by analyzing x-ray diffraction data. We note that this space group involves breaking a rather large number of symmetries of the high-symmetry phase and is not reachable through a continuous (second order) phase transition from the high-symmetry phase (i.e., the symmetry breaking distortion does not transform according to an irreducible representation of $P6_3/mmc$) rendering it energetically more costly~\cite{Franzen74, PerezMato81}. At the same time, the work of Liu et al.~\cite{Liu2023} suggests the presence of the $P6_3mc$ (No. 186) space group, based on HAADF-STEM observations of Fe atoms displacement along the $c$-axis, which depends on Fe content. Indeed, this suggested space group symmetry corresponds to a (maximal) subgroup of the high symmetry $P6_3/mmc$ space group that breaks inversion symmetry and hence would fulfill the prerequisite for hosting N\'eel skyrmions while being compatible with a continuous (Landau-type) phase transition~\cite{PerezMato81, Franzen74}. Such as a higher symmetric non-centrosymmetric phase may exist as a metastable phase or due to inhomogeneous doping and disorder in Fe-deficient systems. In this work, we address the question of whether this minimal inversion symmetry breaking can be directly verified by electron diffraction methods, thereby directly verifying the presence of the $P6_3mc$ space group symmetry.

 To address this question, this work focuses on a direct experimental investigation of the structural symmetry of single crystalline $\text{Fe}_{\text{3-x}}\text{GeTe}_{\text{2}}$ ($\text{x}\approx0.1$) by means of electron diffraction (ED) and convergent beam electron diffraction (CBED) at room-temperature. (CB)ED has been widely utilized to resolve the space groups of various materials in the past since it is capable of distinguishing the presence symmetry operations such as rotational symmetry, mirror symmetry, glide plane, and screw axis directly from observed symmetries in the recorded (CB)ED patterns~\cite{Tanaka11, Morniroli12}, which is in contrast to x-ray or neutron diffraction, where direct probing of inversion symmetry is hampered by Friedel’s law. Moreover, the converged electron beam employed in CBED has a diameter in the range of 10 nm, hence allows probing the symmetry on a local scale (in contrast to x-ray and neutron diffraction probing the average crystal symmetry over macroscopic length scales). E.g., an inspection of the rotational symmetries of CBED patterns recorded along the $[0001]$ zone axis immediately reveals, whether $\text{Fe}_{\text{3-x}}\text{GeTe}_{\text{2}}$ locally belongs to the hexagonal or trigonal system. Similarly, the existence of mirror symmetry along c-direction can be verified from CBED patterns recorded from a $[uvw0]$ zone axis.

\section{Convergent Beam and Selected Area Electron Diffraction}

Electron diffraction experiments, employing two different illumination modes in the TEM, parallel beam and convergent beam, have been conducted to probe the symmetry of $\text{Fe}_{\text{3-x}}\text{GeTe}_{\text{2}}$ single crystals. All experiments have been carried out at a FEI~Titan$^3$~80-300~TEM, equipped with both probe and imaging aberration correction, at 300~kV acceleration voltage.

\subsection{Specimen Preparation}

Single crystals of Fe-deficient $\text{Fe}_{\text{3-x}}\text{GeTe}_{\text{2}}$ ($\text{x}\approx0.1$) have been acquired from HQ graphene, Inc. as in Ref.~\cite{Chakraborty22}. Thin TEM lamellae of zone axis orientation $[10\bar{1}0]$ ($[0001]$) with a thickness of approximately 55~nm (70-85~nm) have been cut via focused Ga$^+$ ion beam milling (employing ThermoFisher Helios 5 CX FIB) after depositing a combined C and Pt protection layer on the surface of the crystal. After lift-out of the lamellae, a low-energy ion beam of 5~kV has been used for final polishing in order to remove surface damaged region. 
Additionally, thin flakes with a thickness of approximately 40~nm that are oriented in $[0001]$ zone axis have been exfoliated with a scotch tape. 

\subsection{SAED and CBED}

The determination of crystal symmetries by electron diffraction techniques has been carried out by employing both parallel illumination electron diffraction (i.e., selected area electron diffraction – SAED) and convergent beam illumination electron diffraction (CBED). We used the SAED patterns to determine the Bravais lattice of $\text{Fe}_{\text{3-x}}\text{GeTe}_{\text{2}}$, where the primitive centering and the presence of the screw axis $6_3$ is read off from systematic absences of reflections. The CBED patterns were recorded on a Gatan OneView camera with an exposure time of approximately $0.1$~s and a probe current around $100$~pA to ensure good signal-to-noise ratios. The convergent beam illumination aperture of $70$~$\mu m$ was chosen in order to obtain large, non-overlapping (in one instance also overlapping) diffraction disks with discernible contrast features, facilitating symmetry determination.

To uniquely characterize the point group symmetries of $\text{Fe}_{\text{3-x}}\text{GeTe}_{\text{2}}$, CBED patterns have been recorded along a sufficient set of zone axis orientations. For hexagonal (e.g., $P6_3/mmc$) and trigonal (e.g., $P3m1$) symmetries, these zone axis orientations are $[0001]$ and $[uvw0]$, where the latter orientation probes the existence of mirror symmetry perpendicular to the c-direction. Hence, the CBED measurements clarify directly, whether the crystal belongs to the hexagonal or trigonal system, with mirror symmetry along $c$-direction or not. The presence or absence of symmetries in the CBED patterns have been analyzed quantitatively by computing the Euclidean difference between original and symmetry transformed CBED patterns. The details of that algorithm are described in Appendix \ref{app:symmetrycomputation}. We furthermore reiterate that CBED is a local probe, i.e., the diameter of the illumination spot of the convergent electron beam in the sample plane is around 10 nanometers. Thus, in spite of having recorded CBED patterns at several positions as well as overview images and parallel ED patterns of large areas that indicate good single crystal quality, we cannot exclude the presence of other crystal symmetries than those discussed in the following.

The datasets discussed in section \ref{sec:res} are available online~\cite{DataSetRadar}.

\section{Results}\label{sec:res}

\subsection{Diffraction along $[0001]$ zone axis}

\begin{figure*}
\centering{\includegraphics[scale = 0.345]{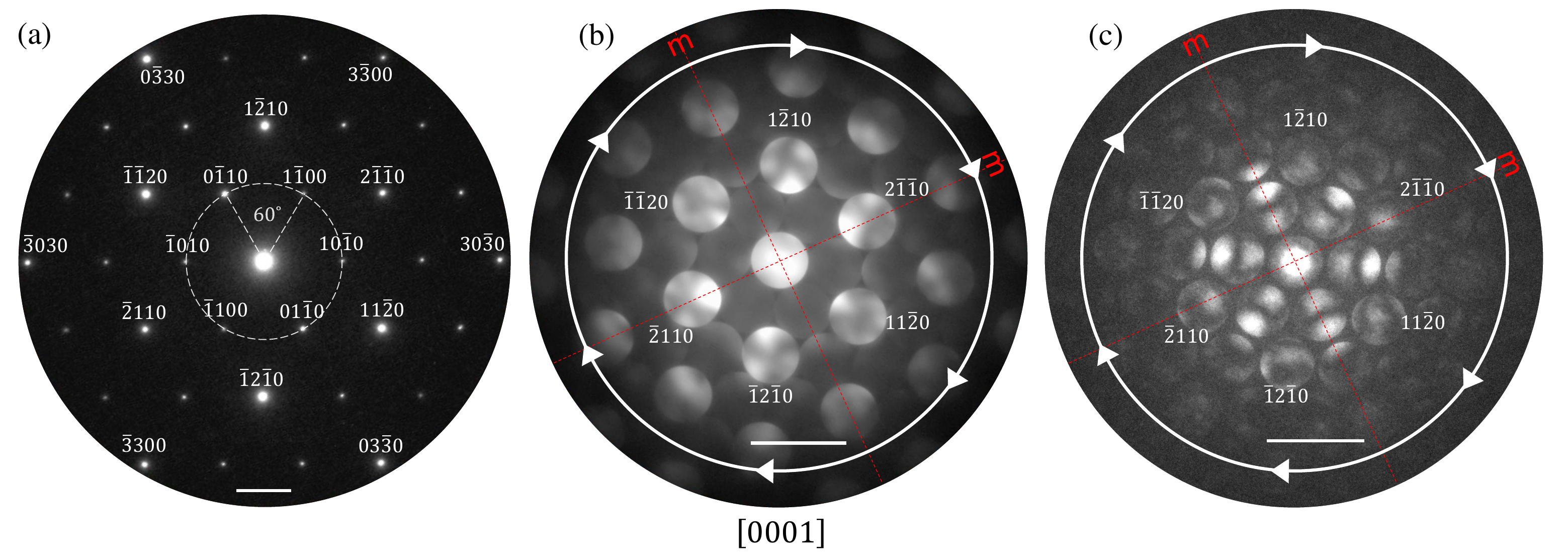}}
\caption{Electron diffraction along $[0001]$ zone axis: (a) selected area diffraction pattern (scale bar is 2~nm$^{-1}$), (b) CBED pattern of exfoliated sample at convergence angle of 3~mrad (scale bar is 5~nm$^{-1}$) and (c) CBED pattern on FIB-cut sample at convergence angle of 2.7~mrad (scale bar is 5~nm$^{-1}$), indexed by Miller-Bravais indices.}
\label{fig:0001experiment}
\end{figure*}

Fig.~\ref{fig:0001experiment} shows SAED (a) and CBED (b,c) patterns of exfoliated (a,b) and FIB-cut (c) $\text{Fe}_{\text{3-x}}\text{GeTe}_{\text{2}}$ recorded at room temperature with an electron beam incident along the $[0001]$ direction. All the reflection spots could be indexed with Miller-Bravais indices, corresponding to lattice constants $a \approx 4.0$~\r{A}. The presence of a hexagonal reciprocal lattice establishes presence of a 3-fold rotational symmetry at this point. Sharpness and symmetry of the SAED patterns indicate high crystallinity and homogeneity of the original crystal. An additional investigation of higher order Laue zone symmetries (HOLZ) in Appendix~\ref{app:HOLZ} confirms the primitive hexagonal lattice of the investigated sample.

To determine whether $\text{Fe}_{\text{3-x}}\text{GeTe}_{\text{2}}$ belongs to either a trigonal or hexagonal system, we studied the symmetries of CBED patterns (Fig.~\ref{fig:0001experiment}b,c). Here, the quantitative symmetry analysis (Appendix \ref{app:symmetrycomputation}) of the CBED patterns reveal a decrease of the $R$ value quantifying the presence of symmetries (the smaller, the better) from 3-fold to 6-fold, $y$-mirror and $x$-mirror symmetries (see Tab.~\ref{table:CBEDquant}). Taking into account the previously established presence of the 3-fold symmetry by SAED, the lower $R$ values of 6-fold, $y$-mirror, and $x$-mirror symmetries suggest that they are also present, which corresponds to a CBED diffraction group symmetry of $6mm$. The small but noticeable breaks of these symmetries in the CBED patterns are attributed to surface damage during sample preparation as well as surface adsorbates.

\subsection{Diffraction along $[10\bar{1}0]$ zone axis}

\begin{figure*}
\centering{\includegraphics[scale = 0.5]{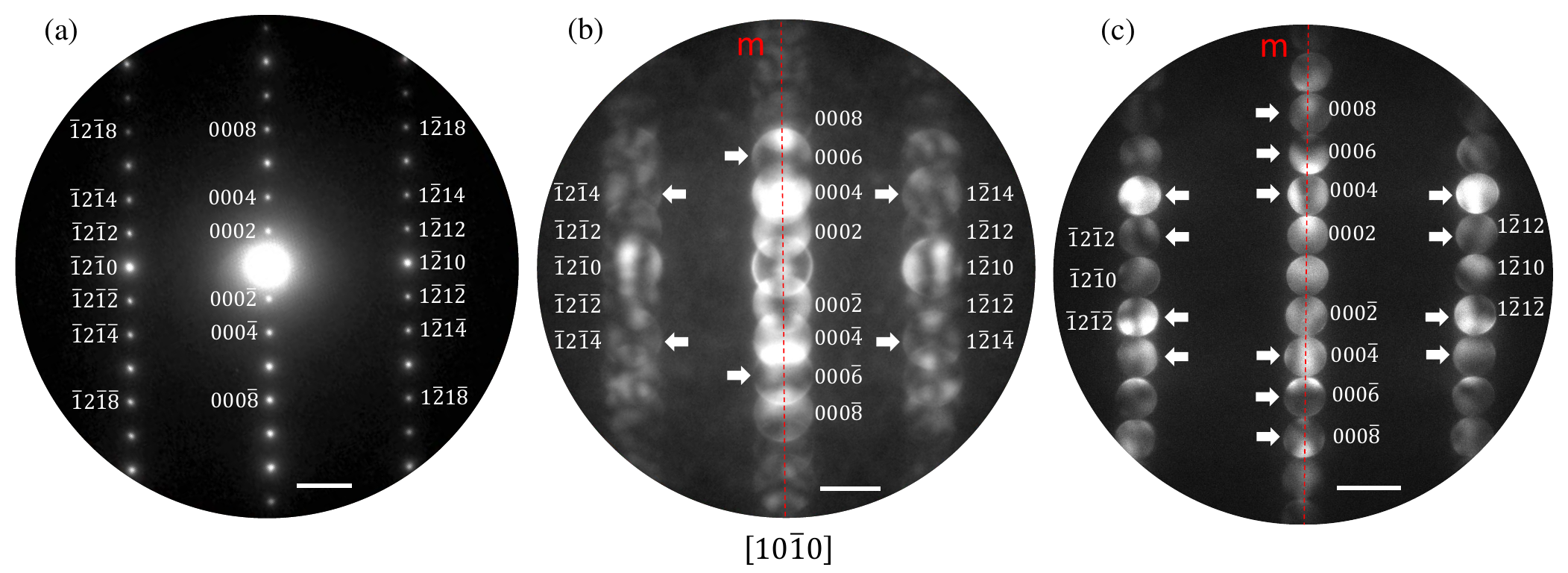}}
\caption{Electron diffraction along $[10\bar{1}0]$ zone axis : (a) selected area diffraction pattern (scale bar is 2~nm$^{-1}$), (b) CBED pattern at convergence angle of 2~mrad (scale bar is 2~nm$^{-1}$), and (c) another CBED pattern at  smaller convergence angle of 1.25~mrad recorded at a different position both suggesting presence of a vertical mirror symmetry (scale bar is 2~nm$^{-1}$).}
\label{fig:10-10experiment}
\end{figure*}

Fig.~\ref{fig:10-10experiment} shows SAED (Fig.~\ref{fig:10-10experiment}a) and CBED (Fig.~\ref{fig:10-10experiment}b,c) patterns of $\text{Fe}_{\text{3-x}}\text{GeTe}_{\text{2}}$ at room temperature recorded along $[10\bar{1}0]$ electron beam incidence. All the reflection spots could be indexed with Miller-Bravais indices with lattice constants $a \approx 4.0$~\r{A} and $c \approx 16.2$~\r{A}. The diffraction pattern exhibits systematic absences, such as 000l=odd, which are consistent with the presence of a $6_3$ screw axis and $c$ glide plane. Additional analysis of forbidden reflections for $[11\bar{2}0]$ zone axis in Appendix~\ref{app:HOLZ} further confirms this result. Sharp, symmetrical peaks in SAED patterns again indicate high crystallinity and homogeneity of the original crystal.

Fig.~\ref{fig:10-10experiment}b and c shows the CBED patterns of $\text{Fe}_{\text{3-x}}\text{GeTe}_{\text{2}}$ at room temperature recorded along $[10\bar{1}0]$ zone axis at different locations and different convergence angles. The symmetry quantification of the $[10\bar{1}0]$ CBED patterns suggests the presence of one mirror symmetry (Tab. \ref{table:CBEDquant}). Both a horizontal mirror plane and a 2-fold rotational symmetry cannot be fitted with $R$ values lower than 0.45 due to sizable differences in the intensity distribution within the CBED discs (e.g., indicated by white arrows in Figs.~\ref{fig:10-10experiment}b).

\begin{table}
\begin{center}
\begin{ruledtabular}
\begin{tabular}{ccc}
 zone axis & symmetry & $R$ \\ 
 \hline
 $\left[0001\right]$ exfoliated & 3 & 0.19 \\ 
 $\left[0001\right]$ exfoliated & 6 & 0.17 \\
 $\left[0001\right]$ exfoliated & m$_x$ & 0.15 \\
 $\left[0001\right]$ exfoliated & m$_y$ & 0.12 \\
 $\left[0001\right]$ FIB & 3 & 0.24 \\ 
 $\left[0001\right]$ FIB & 6 & 0.2 \\
 $\left[0001\right]$ FIB & m$_x$ & 0.23 \\
 $\left[0001\right]$ FIB & m$_y$ & 0.23 \\
 $\left[10\bar{1}0\right]$ & m$_x$ & 0.21 \\ 
 \end{tabular}
\end{ruledtabular}
\end{center}
\caption{Symmetry quantification results for experimental CBED patterns displayed in Figs. \ref{fig:0001experiment} and \ref{fig:10-10experiment}.}
\label{table:CBEDquant}
\end{table}

\section{Discussion}

The SAED and CBED patterns obtained at [0001] and [10$\bar1$0] electron beam incidence reveal diffraction group symmetries consistent with $6mm$ and $m$, respectively. From that, the point group of $\text{Fe}_{\text{3-x}}\text{GeTe}_{\text{2}}$ at room temperature can be fixed to $6mm$ upon inspection of coincidence tables~\cite{Buxton76}. That also entails that the Bravais lattice is hexagonal, as previously determined by SAED and HOLZ reflections analysis (see Appendix~\ref{app:HOLZ}). To identify the corresponding space group symmetry, we first note the presence of the $6_3$ screw axis, determined from systematic absences in the SAED (Fig.~\ref{fig:10-10experiment}a and Fig.~\ref{fig:ForbRef11-20}). The presence of the $6_3$ screw axis narrows the possible choice of space groups pertaining to the $6mm$ point group to either $P6_3cm$ (No.185) or $P6_3mc$ (No.186), with the latter being the natural choice considering the high-symmetry space group $P6_3/mmc$ of the parent compound $\text{Fe}_{\text{3}}\text{GeTe}_{\text{2}}$ and forbidden reflections analysis (see Appendix~\ref{app:HOLZ}). Indeed, $P6_3mc$ derives from $P6_3/mmc$ by removing the mirror plane perpendicular to the rotational symmetry axis, i.e., it is a maximal subgroup with index 2 of $P6_3/mmc$. Moreover, this subgroup is reachable from $P6_3/mmc$ by a continuous phase transition, i.e., at minimal energy cost \cite{PerezMato81}. Therefore, the local space group of $\text{Fe}_{\text{3-x}}\text{GeTe}_{\text{2}}$ at room temperature can be identified with $P6_3mc$ (No.186).

\begin{figure*}
\centering{\includegraphics[scale = 0.405]{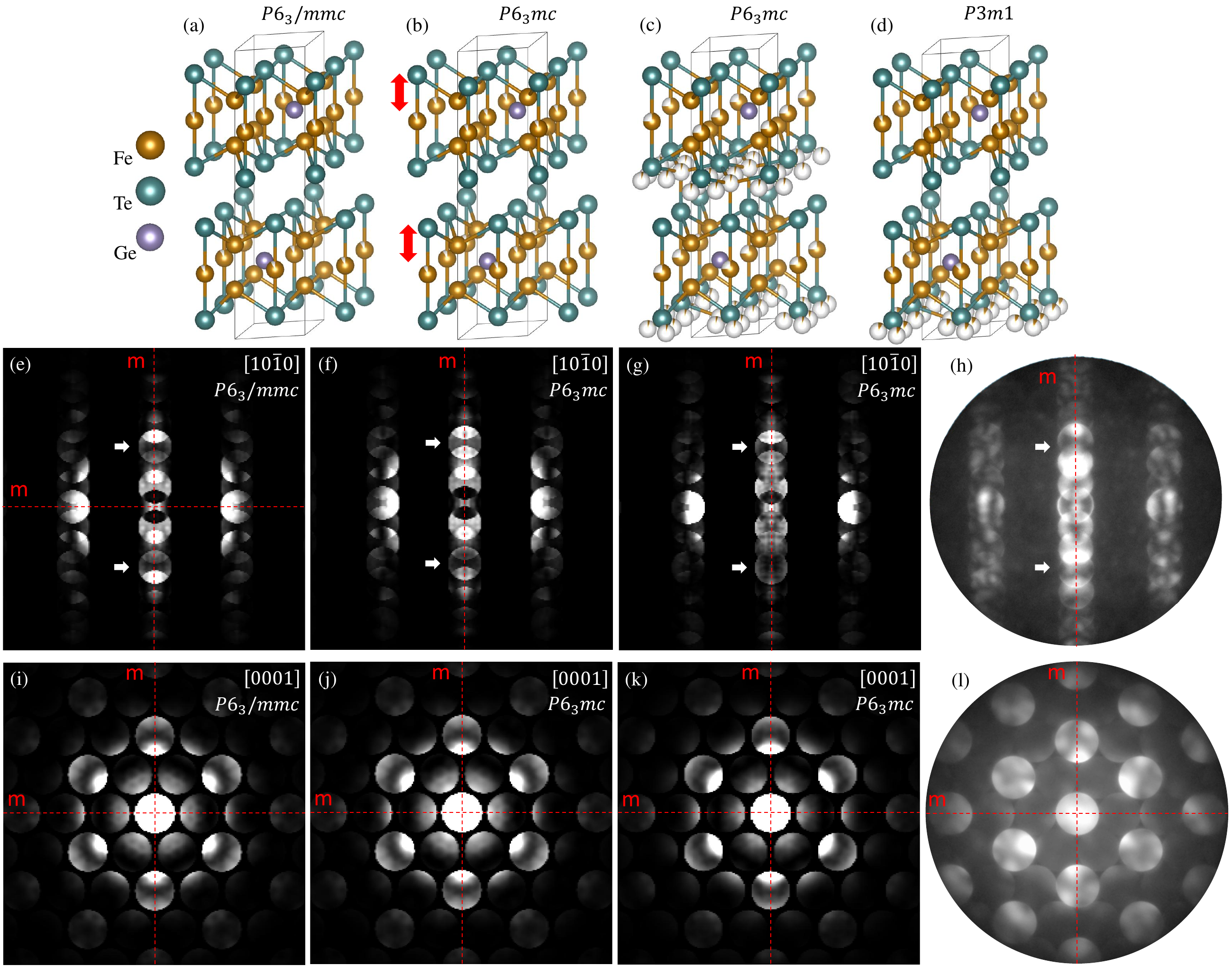}}
\caption{Analysis of possible symmetry differences in $\text{Fe}_{\text{3-x}}\text{GeTe}_{\text{2}}$.  The first row shows (a) parent $P6_3/mmc$ symmetry of $\text{Fe}_{\text{3-x}}\text{GeTe}_{\text{2}}$ \cite{Verchenko15}, (b) possible continuous transformation into $P6_3mc$ symmetry via small shifts of atomic layers along c-axis (indicated by red arrows), (c) transformation into $P6_3mc$ via introduction of Fe atoms inside the vdW’s gaps, and (d) reduction of symmetry down to $P3m1$ via introduction of Fe atoms inside the vdW’s gap proposed by \cite{Chakraborty22}. The second row shows results of CBED simulations for $[10\bar{1}0]$ zone axis with thickness of 54~nm and convergence angle of 2~mrad for (e) original $P6_3mmc$ space group, (f) $P6_3mc$ space group through small ($O$(pm)) continuous transformation of the unit cell, (g) $P6_3mc$ space group through occupation of different sites within the vdW’s gap by Fe atoms, and (h) experimental $[10\bar{1}0]$ CBED pattern, previously  shown in Fig.~\ref{fig:10-10experiment}. The third row shows results of CBED simulations for $[0001]$ zone axis with thickness of 36.5~nm and convergence angle of 2.5~mrad for (i) original $P6_3mmc$ space group, (j) $P6_3mc$ space group through small ($O$(pm)) continuous transformation of the unit cell, (k) $P6_3mc$ space group through occupation of different sites within the vdW’s gap by Fe atoms, and (l) experimental $[0001]$ CBED pattern, previously shown in Fig.~\ref{fig:0001experiment}. The experimental patterns have been recorded without energy filtering, which explains the significantly stronger background.}
\label{fig:simul}
\end{figure*}

To shed further light on the symmetry reduction to $P6_3mc$ including possible symmetry breaking structural modifications of the high-symmetry parent compound we carried out CBED simulations (see Fig.~\ref{fig:simul}) employing the Dr. Probe software \cite{Barthel18}. For the simulations 300~kV acceleration voltage and a convergence angle of 2 and 2.5~mrad similar to the experiments have been used. Two specific $P6_3mc$ crystal structures were simulated to explore possible symmetry breaking mechanisms. The first structure displayed in Fig.~\ref{fig:simul}~b was derived from the high-symmetry parent compound (Fig.~\ref{fig:simul}~a) by slightly shifting ($O$(pm)) the Fe atoms along the $c$-direction, thereby breaking the horizontal mirror symmetry, while preserving the other symmetries. This is an example of possible continuous transformation. The second $P6_3mc$ structure in Fig.~\ref{fig:simul}~c is obtained via a horizontal-mirror-symmetry breaking introduction of Fe atoms inside all vdW’s gaps in accordance with the Ref. \cite{Chakraborty22}, where a selective filling of every second vdW’s gap has been proposed to reduce the symmetry even further down to $P3m1$. At a sample thickness of approximately 36.5~nm, the simulated $[0001]$ CBED patterns of both structures (Fig.~\ref{fig:simul}~i and j) are in good agreement with the experimental CBED pattern (Fig.~\ref{fig:0001experiment}b). The comparison of the $[10\bar{1}0]$ zone axis CBED patterns (Fig.~\ref{fig:10-10experiment}b) yields a better agreement with the simulated CBED patterns of the first structure at a slab thickness of 54~nm (Fig.~\ref{fig:simul}~f) than with the second structure (Fig.~\ref{fig:simul}~g). 

For comparison, we also show simulated CBED patterns of the high-symmetry $P6_3/mmc$ (Fig.~\ref{fig:simul}a) parent compound~\cite{Verchenko15}, exhibiting $6mm$ CBED symmetry along $[0001]$ zone axis (Fig.~\ref{fig:simul}h) and $2mm$ CBED symmetry along $[0001]$ zone axis (Fig.~\ref{fig:simul}e), with the horizontal mirror symmetry not present in the experiment (Fig.~\ref{fig:10-10experiment}).

As the first structure in Fig.~\ref{fig:simul}~b can be obtained from the $P6_3/mmc$ parent structure (Fig.~\ref{fig:simul}~a) by a continuous phase transition consisting of a slight shift of atoms at minimal energy cost \cite{PerezMato81}, we conclude that this mechanism is the more likely mechanism for the symmetry reduction to $P6_3mc$. Additionally, the recent work of Liu et al.~\cite{Liu2023} revealed, using HAADF-STEM, the displacement of Fe atoms along the $c$-axis depending on the Fe content, that would be consistent with a symmetry reduction to $P6_3mc$. To fully confirm this, however, an XRD or ED refinement of the structure would be necessary, which is beyond the scope of this paper.

\section{Conclusion}

Using a combination of SAED and CBED, we found the point group and space group symmetry of $6mm$ and $P6_3mc$ (No.186), respectively, in $\text{Fe}_{\text{3-x}}\text{GeTe}_{\text{2}}$ at room temperature. The unit cell dimensions are similar to the previously reported $P3m1$ (No.156) and the high-symmetry parent compound $\text{Fe}_{\text{3}}\text{GeTe}_{\text{2}}$ with space group $P6_3/mmc$ (No.194). In contrast to $P3m1$, $P6_3mc$ is reachable from the parent compound $P6_3/mmc$ through a continuous structural phase transition. Similar to $P3m1$, the absence of centrosymmetry in the $P6_3mc$ phase would support the formation of N\'eel skyrmions and bubbles in $\text{Fe}_{\text{3-x}}\text{GeTe}_{\text{2}}$ as revealed in previous studies.  As the three space groups - $P6_3/mmc$, $P6_3mc$ and $P3m1$ - form a chain of subgroups, our study indicates a scenario, where changes in the stoichiometry (e.g., Fe deficiency) lead to sequential removal of symmetries, with the mirror symmetry perpendicular to the 6-fold axis being broken first and the screw and glide plane symmetry second. Future studies may reveal the precise crystal structure, e.g., with respect to preferred positions and filling of Fe sublattice, clarifying the symmetry breaking mechanism.

\section{Acknowledgements}

This work was supported by the Deutsche Forschungsgemeinschaft through SFB 1415, project-id 417590517. This work also received support from the Deutsche Forschungsgemeinschaft DFG through the SFB 1143, project-id 24731007.

\appendix

\section{Quantitative symmetry analysis of experimental CBED patterns} 
\label{app:symmetrycomputation}

\begin{figure}
\centering{\includegraphics[width = \columnwidth]{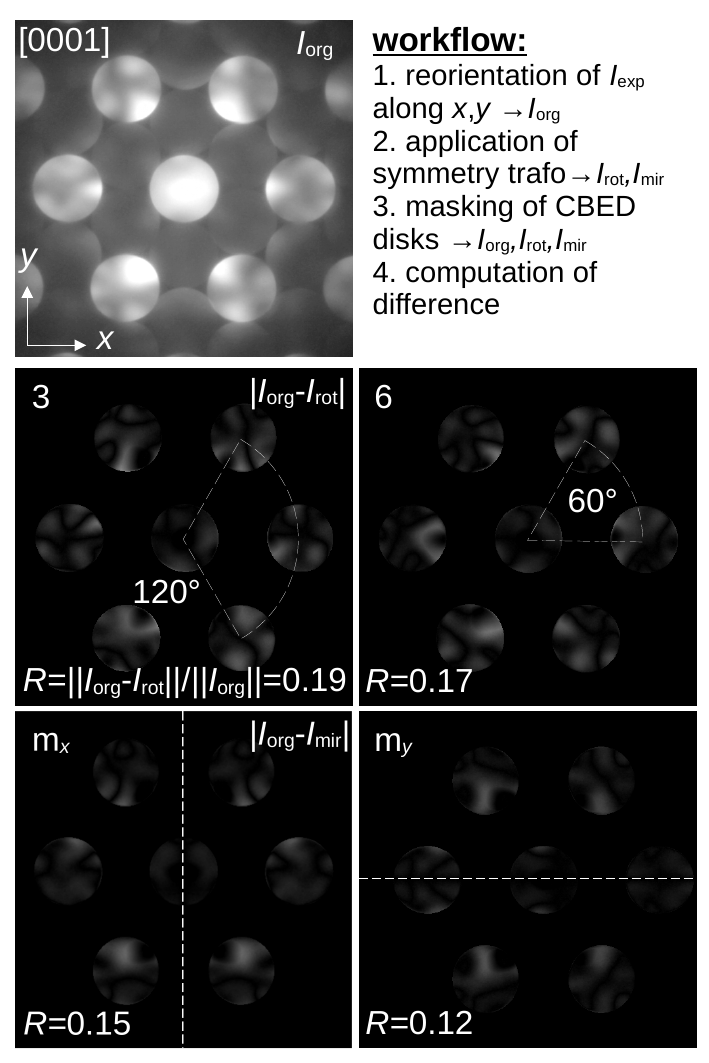}}
\caption{Symmetry analysis of $[0001]$  CBED patterns. In a first step the original CBED pattern (Fig. \ref{fig:0001experiment}(b) was aligned to the $xy$ coordinate system. Subsequently, the rotational and mirror symmetries have been applied, respectively. The CBED discs have been masked before computing the normalized difference between the original and transformed CBED discs. The differences are plotted in the same gray scale as the original CBED pattern to intuitively highlight the magnitude of the differences.}
\label{fig:001symanalysis}
\end{figure}

To quantitatively assess the presence or absence of symmetries in the experimental CBED patterns we developed an algorithm that computes the Euclidean difference between original and the symmetry transformed CBED pattern also taking into account the effect of residual misorientation between $[0001]$ and $[10\bar{1}0]$ zone axis. The principal steps of the algorithm are
\begin{enumerate}
    \item Rotational alignment of the original CBED pattern with a Cartesian coordinate system that is aligned with the possible mirror planes of the pattern.
    \item Application of symmetry transformation to CBED pattern (either rotational or mirror).
    \item Masking of a predefined set of CBED disks in order to remove the impact of background in between the disks.
    \item Computation of Euclidean distance between original and symmetry-transformed masked CBED pattern $R=\frac{\parallel I_\mathrm{org} - I_\mathrm{trafo}\parallel}{\parallel I_\mathrm{org} \parallel}$ 
\end{enumerate}

These steps are iterated, while varying the origin of the symmetry transformation (i.e., the invariant point or line) thereby taking into account the effect of residual beam tilt or crystal tilt that effectively “shifts” the CBED pattern under the aperture given by the convergence angle. In the present case only small residual shifts were observed, which were not responsible for the observed magnitude of the $R$ values. The algorithm is implemented as a python script plugin in the Panta Rhei image processing software of CEOS \cite{PantaRhei}. The algorithm is available online~\cite{CBEDquant}.

\begin{figure}
\centering{\includegraphics[width = \columnwidth]{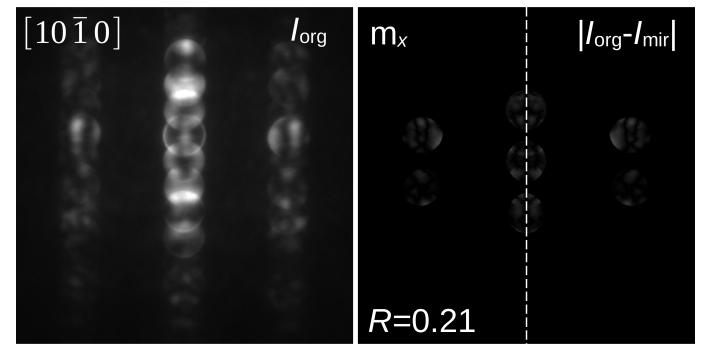}}
\caption{Symmetry analysis of $[10\bar{1}0]$  CBED patterns. In a first step the original CBED pattern (Fig. \ref{fig:10-10experiment}(b) was aligned to the $xy$ coordinate axis. Subsequently, the rotational and mirror symmetries have been applied respectively. The CBED discs have been masked before computing the normalized (Euclidean) difference between the original and transformed CBED discs.}
\label{fig:210symanalysis}
\end{figure}

\section{Analysis of HOLZ symmetries and additional forbidden reflections analysis} 
\label{app:HOLZ}

In addition to the analysis of CBED patterns and kinematically forbidden reflections, we recorded and analyzed HOLZ reflections in diffraction patterns, which can provide important additional information about the investigated structure even without perfect alignment with a zone axis or the use of methods such as Precession Electron Diffraction \cite{Morniroli1992,Morniroli12}.

\begin{figure*}
\centering{\includegraphics[scale = 0.51]{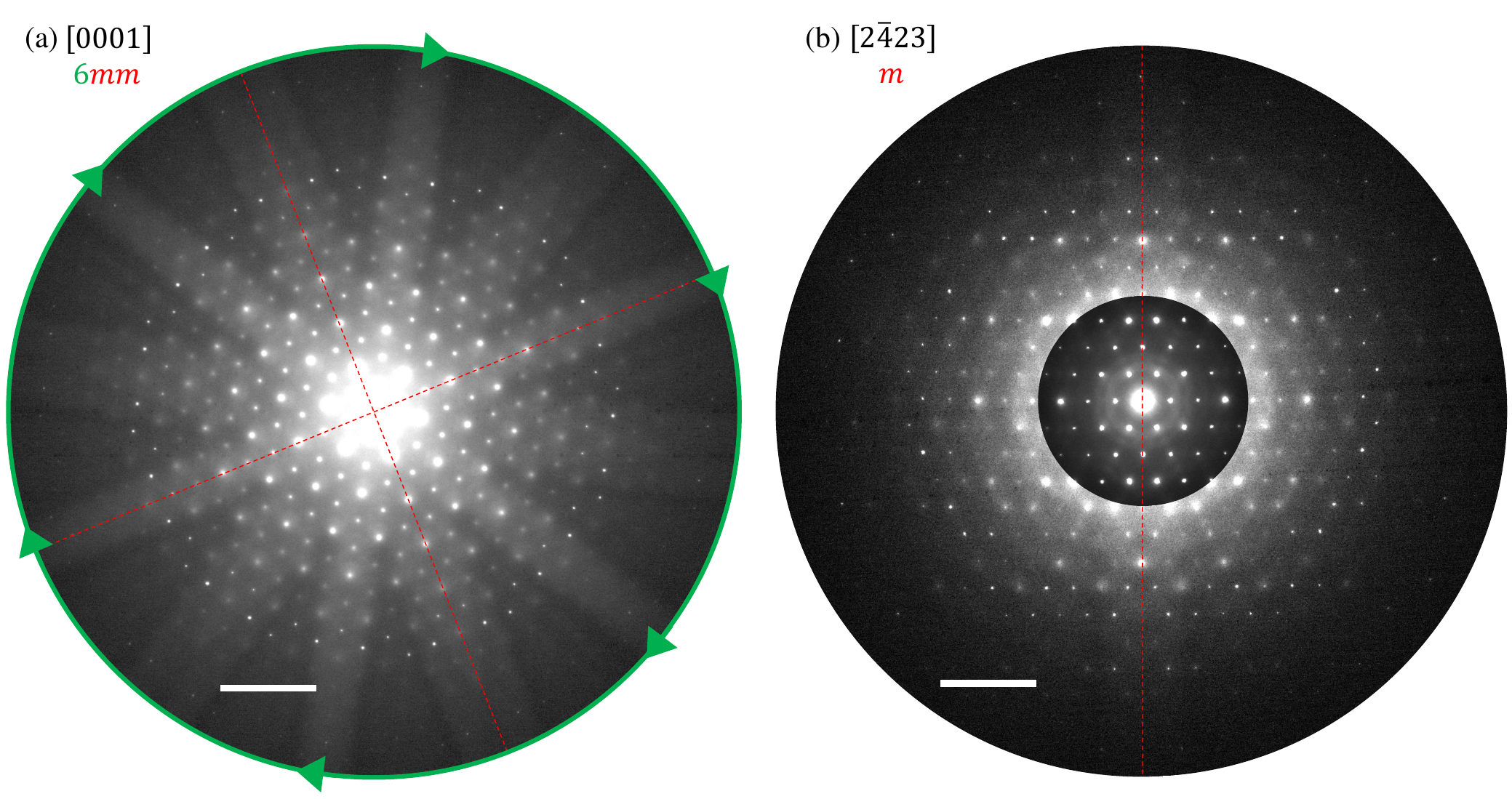}}
\caption{Symmetry analysis of HOLZ patterns: (a) electron diffraction along $[0001]$ zone axis, showing $6mm$ whole pattern symmetry, and (b) electron diffraction along $[2\bar{4}23]$ zone axis, showing $m$ whole pattern symmetry (different contrast within the ZOLZ was used to make reflections visible). Scale bar for both images is 10~nm$^{-1}$.}
\label{fig:HOLZ0001}
\end{figure*}

As a result, whole diffraction pattern symmetries (including HOLZ), recorded along $[0001]$ and $[u\bar{2u}uw]$ zone axis, can be used to determine whether the investigated structure has a rhombohedral centering or a primitive hexagonal one \cite{Morniroli1992}. In our case, the experimental diffraction patterns, recorded along the $[0001]$ (see Fig.~\ref{fig:HOLZ0001}~a) and the $[2\bar{4}23]$ zone axis (see Fig.~\ref{fig:HOLZ0001}~b), show $6mm$ and $m$ whole pattern symmetries, respectively. According to Morniroli and Steeds \cite{Morniroli1992}, these symmetries correspond to a hexagonal lattice. Consequently, we can exclude all rhombohedral space group symmetries, such as $R3c$, $R\bar{3}c$, etc.

\begin{figure*}
\centering{\includegraphics[scale = 0.51]{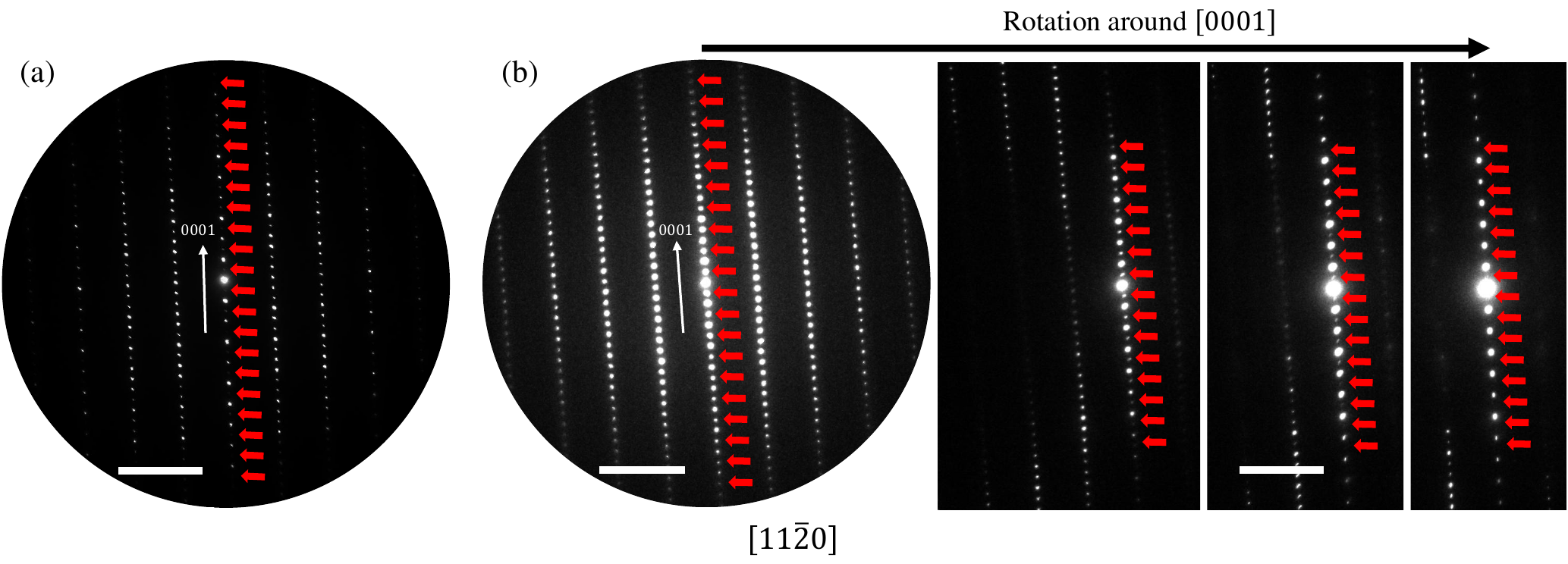}}
\caption{Analysis of kinematically forbidden reflections for the $[11\bar{2}0]$ zone axis: (a) SAED pattern showing directly visible forbidden reflections and (b) SAED pattern showing forbidden reflections allowed by dynamical diffraction effects, which weaken when rotating the sample around the $0001$ axis. Forbidden reflections are indicated by red arrows. The scale bars are 5~nm$^{-1}$.}
\label{fig:ForbRef11-20}
\end{figure*}

The $6_3$ screw axis in combination with a $c$ glide plane should produce $000l$ forbidden reflections with $l = \pm1, \pm3, \dots$ for the $[11\bar{2}0]$ zone axis \cite{Morniroli1992}. In our case, the $[11\bar{2}0]$ diffraction patterns at some locations show forbidden reflections directly (see Fig.~\ref{fig:ForbRef11-20}~a), while for other locations forbidden reflections can appear due to dynamical diffraction effects (see Fig.~\ref{fig:ForbRef11-20}~b). In such cases, we rotate around the $0001$ reciprocal row to confirm that forbidden reflections are present at those locations as well, as they vanish with sufficient rotation \cite{Morniroli1992}. Together with systematic absences for the $[10\bar{1}0]$ zone axis, such as $uvkl$ with $l = \pm1, \pm3, \dots$ (see Fig.~\ref{fig:10-10experiment}~a), these results are consistent with the presence of a $6_3$ screw axis and a $c$ glide plane ($P6_3\dots c$ in particular \cite{Morniroli1992}).

Analysis of CBED symmetries in the main text, combined with HOLZ and forbidden reflections analysis, confirms the presence of the $P6_3mc$ phase.

\bibliography{literature}

\end{document}